\documentstyle[psfig,preprint,aps]{revtex}

\tightenlines

\newcommand{\sla}{\!\!\!\!/ \,}

\def\beq{\begin{equation}}
\def\eeq{\end{equation}}
\def\bea{\begin{eqnarray}}
\def\eea{\end{eqnarray}}

\begin{document}
\hoffset-1cm
\draft

\title{Leontovich Relations in Thermal Field Theory}

\author{Markus H. Thoma\footnote{Heisenberg fellow}}
\address{Theory Division, CERN, CH-1211 Geneva 23, Switzerland}

\date{\today}

\maketitle

\begin{abstract}

The application of generalized Kramers-Kronig relations, the so-called 
Leontovich relations, to thermal field theory is discussed. Medium effects 
contained in the full, thermal propagators can easily be taken into account
by this method. As examples the collisional energy loss of a charged 
particle in a relativistic plasma and the radiation of energetic photons from 
a quark-gluon plasma are considered. Within the leading logarithmic 
approximation the results based on the hard thermal loop resummation
technique are reproduced easily. However, the method presented here
is more general and provides exact expressions, which allow in 
principle non-perturbative calculations.   
\end{abstract} 

\bigskip


\narrowtext

\section{Introduction}

Naively one expects that high energy particles, weakly interacting 
with a medium, can be treated 
perturbatively. The basic process (production, absorption, or scattering) 
should follow from
lowest order perturbation theory as in vacuum. The only
role of the medium is to
provide the particles with which the high energy particle interacts.
In other words, medium effects enter the cross section only via the 
distribution functions of the in-medium particles. Note, however, that in the 
case of bosons the cross sections can be infrared enhanced due to Bose 
condensation. Famous examples are
the effective medium dependent masses of neutrinos interacting with
the solar or terrestrial matter leading to medium induced neutrino
oscillations. These masses follow 
directly from integrating over the electron distribution \cite{Not88}.
Note that it is not even necessary that these distributions are in 
equilibrium. Another example is the production of dileptons with high 
invariant masses from a quark-gluon plasma (QGP), which is given to lowest 
order by the annihilation of bare quarks and anti-quarks (Born term)
\cite{Cle87}. Due to phase space the main contribution comes from quarks 
in the high energy tail of the distributions. At lower invariant 
masses ($M<1$ GeV) bare quarks are not sufficient and additional medium
effects will lead to interesting structures in the dilepton production 
rate \cite{Pes00}. 

The quantities mentioned above, namely the effective neutrino mass
and the dilepton production, are infrared finite to lowest order 
perturbation theory. In some cases, however, the lowest order contribution 
suffers from infrared divergences. In this case
additional medium effects even for high energy, 
weakly interacting particles are essential. If the quantity under 
consideration has a logarithmic infrared singularity within
naive perturbation theory, a finite result can be obtained by using hard 
thermal loop (HTL) resummed propagators for soft momentum transfers
\cite{Bra91}. However, some quantities, e.g. damping rates \cite{Pis89},
which exhibit a higher infrared singularity, cannot be calculated to leading 
order using the HTL method. Also contributions beyond the leading
logarithm might require a non-perturbative treatment \cite{Aur99}.

Here we will demonstrate, considering two examples,
that the leading logarithmic contributions can be 
calculated easily using generalized Kramers-Kronig relations. In contrast
to the HTL improved perturbation theory the knowledge of the resummed 
propagator is required only in the high frequency limit. The method
presented here is much more general as exact expressions can be derived,
which allow in principle non-perturbative results, if only the full propagator
in the high frequency limit is known.

In the next section we discuss the usefulness of the Leontovich 
relation for thermal field theoretic calculations
in the case of the collisional energy loss of charged particles 
in a relativistic plasma. This quantity has already been considered
by Kirzhnits \cite{Kir87} using the Leontovich method within
the language of plasma physics and recently by the author 
within thermal field theory \cite{Tho00}. Therefore we will discuss
this quantity only briefly focusing on the use of the Leontovich relation.
In section 3 we show in more detail that the production rate of
high energy photons from a quark QGP can be treated in a similar way.
For this purpose we generalize the Leontovich relation used so far
only for gauge bosons to quarks.  

\section{Collisional Energy Loss}

The energy loss of a fast charged particle in a medium
is a well studied subject \cite{Jac75}. Recently the energy loss
of energetic particles, such as leptons and partons, in relativistic plasmas
has attracted great interest. In relativistic heavy ion collisions the energy 
loss of a high energy quark or gluon coming from primary hard collisions
in the fireball will lead to jet quenching. Jets therefore
serve as a direct probe for the fireball and may provide a 
signature for the quark-gluon plasma
formation \cite{Gyu90}. In Supernovae explosions the energy loss of neutrinos,
having a weak charge,
in the plasma surrounding the stellar core might be an important mechanism
for triggering the explosion \cite{Raf96}.

The total energy loss of a particle in a medium can be decomposed into a
collisional and a radiative contribution. While the first one originates 
from the energy transfer to the medium particles, the latter one is caused by
radiation from the fast particle. Here we want to consider only the 
collisional component. 

The collisional energy loss is defined as the 
energy transferred per unit length from the fast particle to the medium 
in a single collision. It is assumed that the fast particle loses only
a small fraction of its energy in each collision. 
In quantum field theory the collisional energy loss 
is defined as 
\cite{Bra91a}
\beq 
\frac{dE}{dx} = \frac{1}{v}\> \int d\Gamma \> \omega,
\label{eq1}
\eeq
where $v$ is the velocity of the incident particle with energy $E$
and $\omega =E-E'$ the energy transfer to the medium. 
The interaction rate $\Gamma $ is identical with the inverse mean free
path. It can be calculated 
either from the matrix element of the process responsible for the energy 
loss or equivalently from the imaginary part of the self energy $\Sigma $ of 
the particle with four momentum $P=(E,{\bf p})$, and mass $M$, ($p=|{\bf p}|$)
\cite{Bra91a}
\beq
\Gamma (E) = -\frac{1}{2E}\> [1-n_F(E)]\> tr[(P\sla +M)\, Im\, \Sigma(E,p)],
\label{eq2}
\eeq
where $n_F(E)=1/[\exp(E/T)+1]$ is the Fermi distribution in the case
of a fermion propagating through a plasma of temperature $T$.
In the following
we restrict ourselves to electrons or muons with high energies $E\gg T$
in an electron-positron plasma. 

To lowest order the interaction rate is caused by elastic scattering of the
fast lepton off the thermal electrons and positrons via one-photon exchange.
Due to the massless photon this rate is quadratically infrared divergent
in naive perturbation theory and cannot be regulated using a HTL photon
propagator containing Debye screening. The collisional energy loss, on the 
other hand, due to the additional factor $\omega $ in the integrand
of (\ref{eq1}) is 
only logarithmically infrared divergent within naive perturbation
theory and finite within the HTL improved perturbation theory. Such a 
quantity can be calculated by introducing a separation scale $eT\ll k^*\ll T$
for the momentum transfer \cite{Bra91}.
Restricting to the leading logarithmic approximation it is sufficient to 
consider the soft momentum transfer $k<k^*$ only. Since the final result
must be independent of $k^*$, it follows from the soft contribution simply 
by replacing $k^*$ by the maximum momentum transfer. In the soft part of the 
energy loss a dressed propagator, containing medium effects such as Debye 
screening, has to be used to regulate the infrared singularity. The exchange of
a soft collective photon or plasma mode corresponds to the energy loss
by polarization of the medium, also known as Fermi density effect
\cite{Jac75}.

The collisional energy loss, caused by the exchange of a single dressed
photon, follows from a one-loop approximation for $\Sigma $. Here we allow 
for the 
most general photon propagator, indicated by the blob in Fig.1. Then we find
for ultrarelativistic electrons or muons ($v=1$)
\cite{Bra91a}
\beq
\left ({\frac{dE}{dx}}\right )_{soft} 
= \frac{e^2}{4\pi }\> \int_0^{k^*} dk\> k \> \int_{-k}^{k}
d\omega \> \omega \> \left [\rho_l(\omega, k)+\left (1-\frac{\omega^2}
{k^2}\right )\rho_t(\omega, k)\right ],
\label{eq3}
\eeq
where $\rho_{l,t}$ are the spectral functions of the full photon propagator, 
defined as 
\beq
\rho_{l,t}(\omega ,k)=-\frac{1}{\pi}\> Im\, D_{l,t}(\omega ,k).
\label{eq4}
\eeq
The full photon propagator fulfills the Kramers-Kronig relation
\beq
D_{l,t}(k_0,k) = \int_{-\infty}^{\infty} d\omega \> 
\frac{\rho_{l,t}(\omega ,k)}{k_0-\omega+i\varepsilon}.
\label{eq5}
\eeq
At finite temperature the photon propagator has only two independent
components, given in Coulomb gauge by the longitudinal 
and transverse propagators \cite{Kap89} 
\bea
D_l(k_0,k) &=& \frac{1}{k^2-\Pi_l(k_0,k)+i\varepsilon},\nonumber \\
D_t(k_0,k) &=& \frac{1}{k_0^2-k^2-\Pi_t(k_0,k)+i\varepsilon},
\label{eq6}
\eea
where $\Pi_{l,t}$ are the longitudinal and transverse components of the
polarization tensor. It should be noted that the soft collisional energy loss, 
discussed here, follows according to (\ref{eq3})
only from the exchange of one dressed space-like ($\omega^2-k^2<0$)
photon from the particle to the medium.
However, the medium particles may undergo further interactions.
The physical process corresponding to the imaginary part of the self energy
of Fig.1 can be found by using cutting rules. An example is shown in Fig.2.
There is no diagram, where to or more photons are emitted from the
fast particle, as in the case of the radiative energy loss. 

Now we introduce the photon response function
\beq
R(k_0,k)=-k^2\> D_l(k_0,k)+(k_0^2-k^2)\> D_t(k_0,k).
\label{eq7}
\eeq
Making the substitution
$k\rightarrow q=\sqrt{k^2-\omega^2}$, i.e. introducing the magnitude
of the four momentum of the exchanged photon, and using $Im\, R(-\omega)
=-Im\, R(\omega )$, which follows from the general property 
$\rho_{l,t}(-\omega )=-\rho_{l,t}(\omega )$ \cite{Sch92}, we find
\beq
\left ({\frac{dE}{dx}}\right )_{soft} = 
\frac{e^2}{2\pi^2}\> \int_0^{q^*} dq\> q\> \int_{0}^{\infty}
d\omega \> \omega\> \frac{Im\, R(\omega,\sqrt{q^2+\omega^2})}{q^2+\omega^2}
\label{eq8}
\eeq
with $q^*\ll T$. Eq. (\ref{eq8}) agrees with Ref.\cite{Kir87},
which is based on plasma physics arguments
if we replace there $Q^2$ by $e^2/4\pi$. 

The response function $R$ fulfills the following Kramers-Kronig
relation \cite{Kir87}
\beq
R(k_0,k)=\tilde R + \frac{2}{\pi}\> \int_0^\infty d\omega \> \omega\>
\frac{Im\, R(\omega,k)}{\omega^2-k_0^2-i\varepsilon},
\label{eq9}
\eeq
which can be shown to be equivalent to 
(\ref{eq4}), if we use $\rho_{l,t}(-\omega)=-\rho_{l,t}(\omega)$. 
Here $\tilde R=\lim_{k_0 \rightarrow \infty} Re\, R(k_0, k)$. 

The Kramers-Kronig relation (\ref{eq9}) can be generalized by 
making a Lorentz transformation from $\omega$ and ${\bf k}$ to 
$\omega '$ and ${\bf k'}$ given in a system which moves with the 
velocity ${\bf u}$ relative to the initial system. Following the arguments 
in Ref.\cite{Leo61,Dol82}, choosing ${\bf u} \cdot {\bf k} =0$ and
$|{\bf u}|=1$, and utilizing that $R(\omega ,{\bf k})$ depends only
on $k$ in an isotropic and homogeneous medium we obtain the Leontovich 
relation \cite{Kir87}
\beq
R(k_0,\sqrt{k^2+k_0^2})=R_\infty + \frac{2}{\pi}\> \int_0^\infty 
d\omega \> \omega\> \frac{Im\, R(\omega,\sqrt{k^2+\omega^2})}
{\omega^2-k_0^2-i\varepsilon},
\label{eq10}
\eeq
where $R_\infty=\lim_{k_0 \rightarrow \infty} Re\, R(k_0,\sqrt{k^2+k_0^2})$.

The $\omega $-integral 
\beq
I=\frac{2}{\pi}\>\int_{0}^{\infty}
d\omega \> \omega\> \frac{Im\, R(\omega,\sqrt{q^2+\omega^2})}{q^2+\omega^2}
\label{eq11}
\eeq 
appearing in  the energy loss (\ref{eq8}) agrees with the 
integral on the right hand side of the Leontovich relation, if we
replace $k_0$ by $iq$ and $\sqrt{k^2+k_0^2}$ by 0, i.e. $k^2=q^2$,
in (\ref{eq10}). Therefore we can write \cite{Kir87}
\beq
I=R(iq,0)-R_\infty.
\label{eq12}
\eeq

The zero momentum limit of the response function vanishes due to the fact
that there is no preferred direction in the medium at vanishing momentum
\cite{Kir87}. Consequently, all we have to know is the response
function in the high frequency limit $R_\infty $ to find the collisional
energy loss in the leading logarithmic approximation. Defining
\beq
\omega_0^2\equiv \lim_{k_0\rightarrow \infty}\Pi_t(k_0,\sqrt{q^2+k_0^2})
\label{eq13}
\eeq
we obtain from (\ref{eq7})
\beq
R_\infty=-\frac{\omega_0^2}{q^2+\omega_0^2}.
\label{eq14}
\eeq

Using the Kramers-Kronig relation for the transverse dielectric function,
which is related to the transverse polarization tensor \cite{Tho00}, 
Kirzhnits argued \cite{Kir87} that $\omega_0$ is independent of $q$ and
can be considered as the effective thermal mass of the transverse 
high frequency plasma excitations, which is given by $\omega_0^2=e^2 n \langle 
1/\Omega \rangle$ in the relativistic limit. Here $n$ is the 
number density of the medium and $\Omega $ the energy of the plasma particles. 

Combining (\ref{eq14}) with (\ref{eq8}) and replacing  $q^*\gg \omega_0$ by
$q_{max}$, which is proportional to $\sqrt{ET}$ in the relativistic limit
$E\gg \Omega$ \cite{Kir87}, 
we end up with the final result for total collisional energy loss
\beq
\frac{dE}{dx} = \frac{e^2}{4\pi}\> \omega_0^2 \ln 
\frac{q_{max}}{\omega_0}.
\label{eq15}
\eeq
This result is ``exact'' in the sense 
that it is independent of any approximation to the full photon 
propagator. To logarithmic 
accuracy the final result just depends on the parameter $\omega_0$.

To proceed from here we have to make an approximation for $\omega_0$.
Since in the high frequency limit medium effects are small, we 
calculate $\omega_0$ by lowest order perturbation theory. Adopting the
gauge invariant expression for the transverse polarization tensor in
the high temperature limit \cite{Kli82}, we obtain
$\omega_0^2 = 3m_\gamma^2/2$, where the plasma frequency is given 
by $m_\gamma =eT/3$. As expected $\omega_0$ is equivalent to the 
high frequency mass of the transverse photon
$\omega_0^2=e^2n\langle 1/\Omega\rangle$ \cite{Tho00}.
Inserting the high temperature result for $\omega_0$ in (\ref{eq15})
leads to an estimate for the collisional energy loss which agrees
to leading logarithm with the one found in the HTL approximation
\cite{Bra91a}
\beq
{\frac{dE}{dx}} = \frac{e^2}{4\pi}\> \omega_0^2 \left (\ln 
\frac{\sqrt{ET}}{\omega_0}+0.120\right ).
\label{eq16}
\eeq

In contrast to the HTL method we only had to know the transverse
polarization tensor
in the high frequency limit. Moreover, we observe that the HTL result has
already the same form as the exact result (\ref{eq15}), which includes
infinitely many higher order diagrams such as the one in Fig.2.
Assuming that the exact high frequency transverse polarization tensor 
can be approximated by its high temperature limit we find that
the complete collisional energy loss can be estimated by its lowest
order HTL result (\ref{eq16}) and that higher order diagrams can be neglected 
at least within the leading logarithm approximation.

In Ref.\cite{Tho00} this result has been applied to the energy loss 
of energetic partons in a QGP. It has been shown that the radiative
energy loss \cite{Bai00} caused by bremsstrahlung from the fast particle, 
which increases linearly with the distance $L$ over which 
the parton propagates, dominates over the collisional one
for $L>1$ fm. 

Another application of (\ref{eq15}) has been
discussed in Ref.\cite{Kir90} in connection with the neutrino
energy loss in matter. Here the collisional energy loss provides
a reliable estimate of the total energy loss since bremsstrahlung
from the neutrino, i.e. emission of a $Z$ boson, is suppressed by the 
large mass of the gauge boson at least for temperatures below $T\simeq 
100$ GeV.

\section{High Energy Photons}

Now we want to discuss another example, namely the production
of high energy real photons in the QGP, which might also serve as a promising
signature for the QGP formation in relativistic heavy ion collisions 
\cite{Ruu92}. Due the weak interaction of photons with the QGP photons
present as well as jet quenching a direct probe for the hot fireball.
To lowest order the production rate of real photons is given by the
diagrams of Fig.3 (Compton scattering, annihilation with gluon absorption).
Here an intermediate bare quark appears, which can be assumed to be massless
as the bare mass of up and down quarks can be neglected compared to the 
temperature $T$ of the QGP. This leads to a logarithmic infrared divergence in
the production rate, which is regulated by medium effects.

The production rate corresponding to the processes
of Fig.3 can be calculated by using the HTL resummed
quark propagator in the case of a soft quark exchange, i.e. momentum exchange
much smaller than $T$. This corresponds to a one-loop calculation including
a HTL quark propagator, that contains an effective quark mass of the order
$gT$, which cuts off the logarithmic singularity. 
For the hard momentum transfer the tree level
scattering matrix elements convoluted with the parton distribution functions
can be used \cite{Kap91,Bai92,Tra95}. In this way the production rate 
of energetic photons to leading logarithmic order $\alpha_s \ln(1/\alpha_s)$
in the strong coupling constant has been obtained. 
 
Here we will derive the photon rate to leading logarithm, using the Leontovich
relation for the full quark propagator, which allows a more general 
and after all simpler derivation of the rate than applying the HTL
method. 
For this purpose we calculate the photon production rate from the
imaginary part of the polarization tensor according to \cite{Kap91}
\beq 
E\frac{dR}{d^3p}=-\frac{2}{(2\pi)^3}\> \frac{1}{e^{E/T}-1}\> Im\, 
\Pi_\mu^\mu (E).
\label{eq17}
\eeq
This expression is exact to all orders in $\alpha_s$ and to leading 
order in $\alpha$. 

Here we focus only on the soft contribution to the photon rate since
the hard contribution can be calculated perturbatively from Fig.1
restricting to the logarithmic approximation. 
Since the energy of the produced photon is high ($E\gg T$) the 
polarization tensor is given by Fig.4. The blob denotes the full 
non-perturbative quark propagator. Due to kinematics there is only one 
full quark propagator, since the other one has to be hard, and no
vertex correction since the high energy photon resolves the vertex 
completely. By cutting this polarization tensor one observes that 
all processes are taken into account, where the soft quark interacts with 
the medium in all possible ways. For example it can absorb a 
thermal gluon as in Fig.3. But also bremsstrahlung from the thermal
particles and other higher order processes are included.

Now we want to calculate the photon production rate from Fig.4 
for the most general full quark propagator without assuming any
approximation for it. For this purpose, we proceed similarly
as in the case of the collisional energy loss for energetic
charged particles in a plasma. We start from
an exact expression for the imaginary part of the soft polarization tensor in 
the case of two massless quark flavors using $E\gg T$ \cite{Kap91}
\beq
Im\, \Pi_\mu^\mu (E) = \frac{5e^2}{12\pi}\> \int_0^\infty
dk \int_{-k}^k d\omega \> [(k-\omega) \rho_+(\omega ,k) + (k+\omega) 
\rho_-(\omega ,k)]\> \theta (q_c^2-k^2+\omega^2),
\label{eq18}
\eeq
where $q_c\ll T$ is the separation scale between the soft and the 
hard contribution, $\omega $ and $k$ the energy and the magnitude
of the three momentum of the soft quark, and $\rho_\pm $ the spectral 
functions of the full quark propagator $S(K)$ in the helicity representation
\cite{Bra90}
\beq
\rho_\pm(\omega ,k)=-\frac{1}{\pi }\> Im\, \frac{1}{D_\pm (\omega ,k)}
\label{eq19}
\eeq
with ($K=(k_0,{\bf k})$)
\beq
S(K)=\frac{\gamma _0-{\bf \hat k} \cdot \mbox{\boldmath$\gamma$}}{2D_+(k_0,k)} +
\frac{\gamma _0+{\bf \hat k} \cdot \mbox{\boldmath$\gamma$}}{2D_-(k_0,k)}.
\label{eq20}
\eeq

Replacing again
$k$ by $q=\sqrt{k^2-\omega^2}$ in the integral of (\ref{eq18})
we find 
\beq
Im\, \Pi_\mu^\mu (E) = -\frac{5e^2}{12\pi^2}\> \int_0^{q_c} dq\> q
\int_{-\infty}^\infty d\omega \> \frac{\omega}{\omega^2+q^2}\> 
Im\, Q(\omega, \sqrt{\omega^2+q^2}),
\label{eq21}
\eeq
where the quark response function $Q$ is given by
\beq
Q(k_0,k) = \frac{k}{k_0}\left (\frac{k-k_0}{D_+(k_0,k)}+
\frac{k+k_0}{D_-(k_0,k)}\right ).
\label{eq22}
\eeq

This response function fulfills the same Kramers-Kronig relation
as the photon response function (\ref{eq9})
\beq
Q(k_0,k)=\tilde Q + \frac{1}{\pi}\> \int_{-\infty}^{\infty} d\omega \>
\omega \> \frac{Im\, Q(\omega ,k)}{\omega^2-k_0^2-i\epsilon},
\label{eq23}
\eeq
where $\tilde Q=\lim_{k_0\rightarrow \infty} Re\, Q(k_0,k)$. 
This relation is a 
direct consequence of the definition of the spectral functions
\beq
\frac{1}{D_\pm(k_0,k)}=\int_{-\infty}^{\infty} d\omega \> \frac{\rho_\pm
(\omega ,k)}{k_0-\omega+i\varepsilon}
\label{eq24}
\eeq
using $\rho_+(-\omega ,k)=\rho_-(\omega ,k)$ \cite{Wel00}.

The Kramers-Kronig relation (\ref{eq21}) can be generalized 
again by the Lorentz transformation of section 2, from which 
we obtain the Leontovich relation analogously to (\ref{eq10})
\beq
Q(k_0,\sqrt{k_0^2+k^2})=Q_\infty + \frac{1}{\pi}\> \int_{-\infty}^\infty 
d\omega \> \omega\> \frac{Im\, Q(\omega,\sqrt{\omega^2+k^2})}
{\omega^2-k_0^2-i\varepsilon},
\label{eq25}
\eeq
where $Q_\infty=\lim_{k_0 \rightarrow \infty} Re\, Q(k_0,\sqrt{k_0^2+k^2})$.
This relation is more restrictive than the Kramers-Kronig relation
(\ref{eq23}) and will be used to evaluate the photon production in
the following from (\ref{eq21}).

The $\omega $-integral in (\ref{eq21}) agrees with the 
integral on the right hand side of the Leontovich relation, if we
replace again $k_0$ by $iq$ and $\sqrt{k_0^2+k^2}$ by 0, i.e. $k^2=q^2$,
in (\ref{eq25}).  Since $D_+(k_0,k=0)=D_-(k_0,k=0)$ \cite{Pes00,Wel00},
$Q(iq,0)=0$. Hence the imaginary part of the polarization tensor
containing the most general in-medium quark propagator is given by 
the simple expression
\beq
Im\, \Pi_\mu^\mu (E) = \frac{5e^2}{12\pi}\> \int_0^{q_c} dq\> q\> Q_\infty.
\label{eq26}
\eeq

Using the Leontovich relation we were able to express the soft part of the 
photon production rate by an integral over the real part of the
response function only in the high frequency limit just below the light cone. 
This expression is exact as long as we do not assume any approximations
for the response function. The advantage of this method is that we do not
have to know the quark 
response function or the quark propagator over the entire
energy range, but only in the high frequency limit. 

Starting from the most general expression for the full quark propagator
(see e.g. Ref.\cite{Pes00})
\beq
D_\pm(k_0,k)=(-k_0\pm k)\> [1+a(k_0,k)]-b(k_0,k)
\label{eq27}
\eeq
and using that the scalar functions $a$ and $b$ fulfill the following 
inequalities in the high frequency limit, where the medium effects on the
quark propagator become small, $\lim_{k_0\rightarrow \infty} a\ll 1$
and $\lim_{k_0\rightarrow \infty} b\ll k_0$,  we find
\beq
Q_\infty =\lim_{k_0\rightarrow \infty}\> \frac{2b_\infty}{q^2/k_0-2b_\infty}.
\label{eq28}
\eeq
Here $b_\infty = \lim_{k_0 \rightarrow \infty} b(k_0,\sqrt {k_0^2+q^2})$.
To proceed we have to 
make an approximation for the full quark propagator or equivalently for
the response function in order to determine $b_\infty$.

In the high frequency limit the response function should be 
calculable perturbatively, as it is also the case for the photon or 
gluon response function
which is related to the dielectric function of the medium \cite{Kir87,Tho00}. 
For in the high frequency limit the dielectric function has to be close to its 
vacuum value and can be computed therefore perturbatively. The same 
argument holds for the quark response function in the high frequency limit. 
Note that the quark response function to 
lowest order, determined from the one-loop quark self energy, 
is infrared finite. In the high temperature limit \cite{Kli82} the gauge 
invariant result 
\beq
b_\infty =-\frac{m_q^2}{k_0}\; \; \; \; \; \Rightarrow \; \; \; \; \;  
Q_\infty=-\frac{2m_q^2}{q^2+2m_q^2}
\label{eq29}
\eeq
is found. Here $2m_q^2=g^2T^2/3$ is the square of the effective high frequency
quark mass. Combining (\ref{eq26}) with (\ref{eq29}) we get
\beq
Im\, \Pi_\mu^\mu (E) = -\frac{5e^2}{12\pi}\> m_q^2 \ln \frac{q_c^2}{2m_q^2},
\label{eq30}
\eeq
where we assumed $q_c\gg m_q$. This result has also be found
by lowest order HTL perturbation theory, where the factor $1/2$ under the 
logarithm could be derived only numerically \cite{Kap91,Bai92}. 
Using, however, 
the Leontovich relation, where one needs to know the response function 
only in the high frequency limit, this factor, related to the high 
frequency effective quark mass, has been obtained analytically.

Combining the soft
part with the hard part, calculated perturbativeky from Fig.3
in Ref.\cite{Kap91}, we obtain the
final result for the production rate of energetic photons to leading logarithm
\beq 
E\frac{dR}{d^3p}=\frac{5}{18\pi^2}\> \alpha \alpha_s\> T^2\> e^{-E/T}\>
\ln \frac{0.2317 E}{\alpha_s T}.
\label{eq31}
\eeq
Here the separation scale $q_c$ serving as an infrared cutoff for the hard 
part drops out since the hard 
part and the soft part have the same factors in front of the logarithm.
This had to be expected by physical reasons, since the final
result has to be independent of the arbitrary separation scale $q_c$
\cite{Bra91}.

The result (\ref{eq31}) agrees with the lowest order HTL 
contribution \cite{Kap91,Bai92}. Unfortunately, this result is of no
practical relevance, as for physical values of the strong coupling constant,
terms beyond the leading logarithm dominate. 
These contributions come from higher order diagrams, 
i.e. two-loop diagrams of the 
HTL perturbative expansion describing e.g. bremsstrahlung,
which show a strong infrared sensitivity
\cite{Aur98}. They are not included in the soft part of the
polarization tensor using the full quark propagator (\ref{eq26}), 
since they come from the exchange of a hard quark \cite{Aur98}.
For realistic values of the coupling these contributions
even dominate clearly over the lowest order ones, in particular at high 
photon energies $E$ as the two-loop contributions are proportional to $ET$ in 
contrast to the one-loop contribution which is proportional to $T^2$. 
As a matter of fact, presumably
infinitely many diagrams within the HTL resummed perturbative expansion 
contribute to order $\alpha_s$ \cite{Aur99}.
Therefore it would be desirable to extend
the arguments given here also to hard momentum transfer. Maybe a resummation
of all higher order diagrams contributing to order $\alpha_s$ leads to a 
cancellation between these diagrams and a suppression of the bremsstrahlung 
and higher processes in the hard photon production. 

\section{Conclusions}

Lowest order perturbation theory at finite temperature in the weak coupling 
limit works only for quantities, which are infrared finite in naive 
perturbation theory. Examples are effective masses and the production
of dileptons with high invariant masses.

Quantities of energetic particles ($E\gg T$)
which are logarithmically infrared divergent in naive perturbation theory, 
such as the
collisional energy loss or the photon production rate, can be calculated 
within the leading logarithm approximation using the lowest order
HTL improved perturbation theory. To extract the leading logarithm it
is sufficient to consider soft momentum transfers described by HTL propagators.
However, for realistic values of the coupling constants, in particular
in the case of the strong coupling constant, contributions beyond the
leading logarithm become important and can even dominate, as in 
the case of the photon production. For the photon 
production rate these contributions have been shown to be 
non-perturbative, i.e. infinitely many higher order diagrams containing 
HTL propagators and vertices contribute to the same order in the coupling 
constant.

Quantities, such as damping rates, which exhibit a higher degree of
infrared divergence in naive perturbation theory
cannot be treated within the HTL improved perturbation
theory. So many properties of (high energy) particles in a medium cannot be 
calculated perturbatively even in the weak coupling limit. 

Here we presented a method for computing within thermal field theory
quantities of energetic
particles, which are in naive perturbation theory
logarithmically infrared divergent, by using generalized Kramers-Kronig
relations. These so-called Leontovich relations follow from the
usual Kramers-Kronig relations for the thermal propagators performing
a Lorentz transformation. Applying these more restrictive relations
one is able to express the soft part of the quantities under consideration,
such as the collisional energy loss or the photon production, by simple
integrals. For evaluating these integrals one needs only the self energy 
of the high energy particle in the high frequency 
limit just below the light cone. In this way an exact expression, i.e.
independent of any approximation for the full propagator,
for these quantities within the logarithmic approximation is obtained.
Assuming perturbation theory to hold for the high frequency self energies
and using the HTL result for them, the results yielded within the lowest order
HTL improved perturbation theory are reproduced. In contrast to the HTL 
resummation technique the method presented here enables analytical
calculations of the soft contributions. Also it is more general, allowing
in principle non-perturbative results, if only the full self energy in
the high frequency limit is known. Since the application of 
perturbation theory at finite temperature
fails in many cases even in the weak coupling limit, it would be desirable
to have non-perturbative or even exact statements. Therefore it might be 
worthwhile to extend the methods presented here also to hard momentum 
transfers, going beyond the leading logarithm.

\vspace*{1cm}

\centerline{\bf ACKNOWLEDGMENTS}
\vspace*{0.5cm}
The author is grateful to G. Raffelt for drawing his attention to the
paper by D.A. Kirzhnits and for helpful discussions and to the 
Max-Planck-Institut f\"ur Physik (Werner-Heisenberg-Institut) for their 
hospitality.

\begin{figure}

\centerline{\psfig{figure=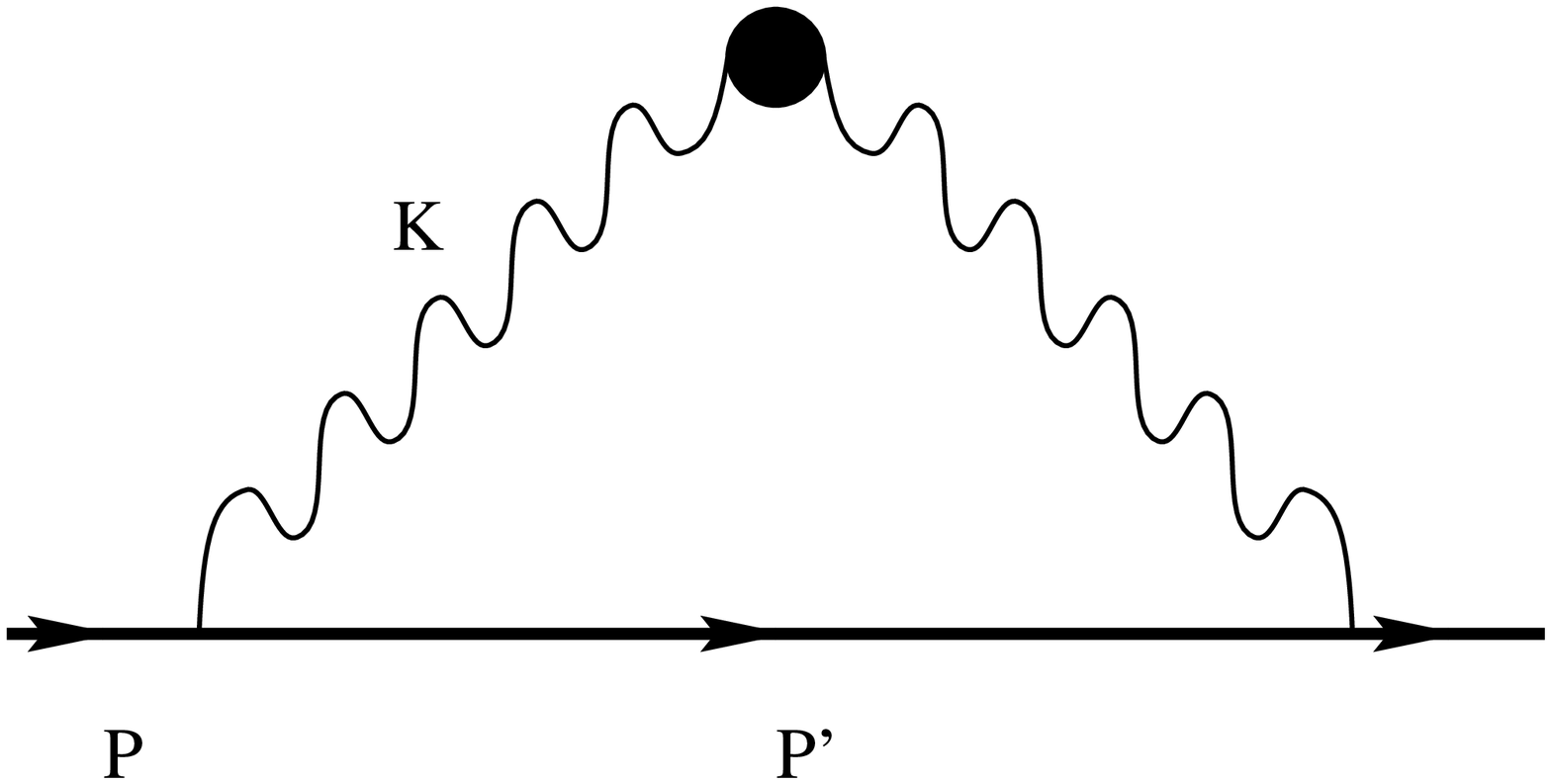,width=8cm}}
\caption{Self energy of a fast fermion containing the full gauge boson 
propagator}

\end{figure}

\begin{figure}

\centerline{\psfig{figure=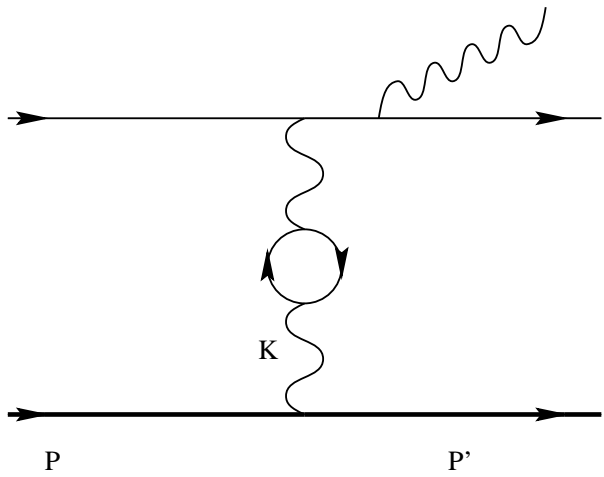,width=8cm}}
\caption{Example for a scattering diagram related to the imaginary part 
of the diagram in Fig.1}

\end{figure}

\begin{figure}

\vspace*{1cm}

\centerline{\psfig{figure=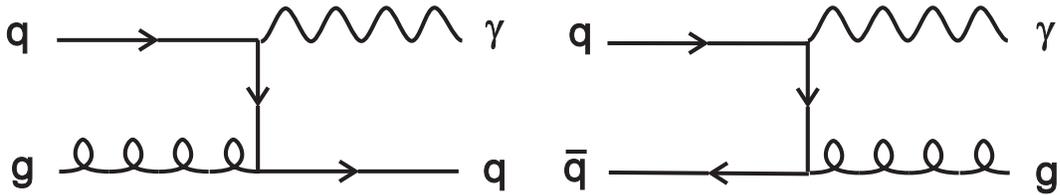,width=14cm}}

\vspace*{1cm}

\caption{Lowest order amplitudes for photon production}

\vspace*{1cm}

\end{figure}

\begin{figure}

\centerline{\psfig{figure=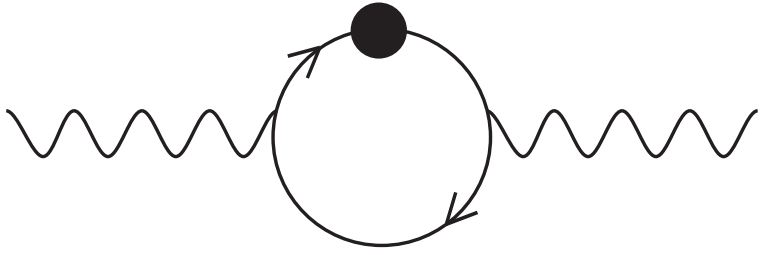,width=10cm}}

\vspace*{0.5cm}

\caption{Polarization tensor containing a full quark propagator}

\end{figure}

\end{document}